\documentclass{elsart3p}

\usepackage{graphicx}

\begin{document}

\newlength{\pixwidth}
\setlength{\pixwidth}{16.5 cm}

\begin{frontmatter}

\title{\boldmath BES3 time of flight monitoring system}

\date{\today}

\maketitle




\author{F. A. Harris\corauthref{cor1}}
\corauth[cor1]{Corresponding author:}
\ead{fah@phys.hawaii.edu}
\author{, J. W. Kennedy, Q. Liu, L. Nguyen, S. L. Olsen,}
\author{M. Rosen, C. P. Shen, G. S. Varner}
\address{Dept. of Physics and Astronomy, University of Hawaii,
Honolulu, Hawaii 96822, U.S.A.}
\author{Y. K. Heng, Z. J. Sun, K. J. Zhu}
\address{Institute of High Energy Physics, C. A. S., Beijing 100049,
People's Republic of China}

\author{Q. An, C. Q. Feng, S. B. Liu}
\address{University of Science and Technology, Hefei 230026, People's
  Republic of China}

\begin{abstract}
  
  A Time of Flight monitoring system has been developed for BES3.
  The light source is a 442-443 nm laser diode, which is stable and
  provides a pulse width as narrow as 50 ps and a peak power as large
  as 2.6 W.  Two optical-fiber bundles with a total of 512 optical
  fibers, including spares, are used to distribute the light pulses to
  the Time of Flight counters.  The design, operation, and performance
  of the system are described.


\end{abstract}


\end{frontmatter}

\section{Introduction}

\begin{figure}  \centering
   \includegraphics*[width=0.50\pixwidth]{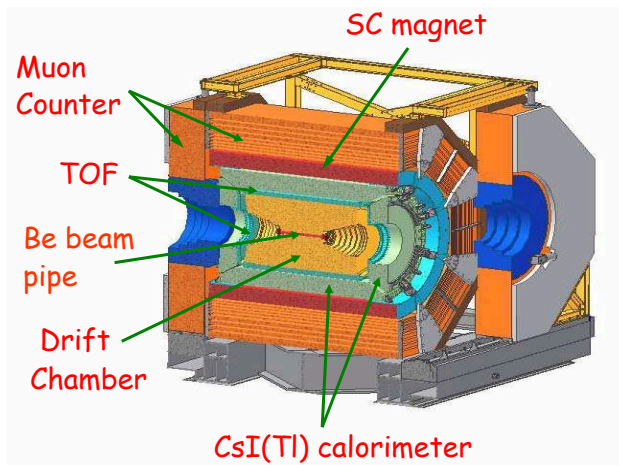}  
  \caption{\label{besiii}BES3 cut-away diagram.
    }
 \end{figure}

The Beijing Electron-Positron Collider (BEPC) and the Beijing
Spectrometer (BES)~\cite{besi,besii} have operated in the tau-charm
center-of-mass energy region from 2 to 5 GeV since 1990.  Currently
they are being upgraded to BEPCII and BES3~\cite{bes3,wli,yifang},
respectively.  BEPCII is a two-ring collider with a design luminosity
of $1 \times 10^{33}$ cm$^{-2}$ s$^{-1}$, an improvement of a factor
of 100 with respect to the BEPC.  BES3, shown in Fig.~\ref{besiii},
is a new detector and features a beryllium beam pipe; a small-cell,
helium-based drift chamber (MDC); a Time-of-Flight (TOF) system; a
CsI(Tl) electromagnetic calorimeter; a 1 Tesla super conducting
solenoidal magnet; and a muon identifier using the magnet yoke
interleaved with Resistive Plate Chambers.  Particle identification is
accomplished in BES3 using $dE/dx$ measurements in the MDC and TOF
measurements.

In this paper, we report on the development and performance of a laser
diode based TOF monitoring/calibration system.  Pulses of light from the
laser diode are injected into one of two optical fiber bundles, one for
each end of the BES3 detector, and the light is delivered by the
individual optical fibers to the TOF counters.  The laser is pulsed under
computer control, and the bundle being illuminated is also under
computer control. The digitized responses from the phototubes mounted
on the TOF counters are then checked and compared with a database.
This allows monitoring of the performance of the TOF system.

\section{Time-of-Flight System}

\begin{figure}  \centering
   \includegraphics*[width=0.46\pixwidth]{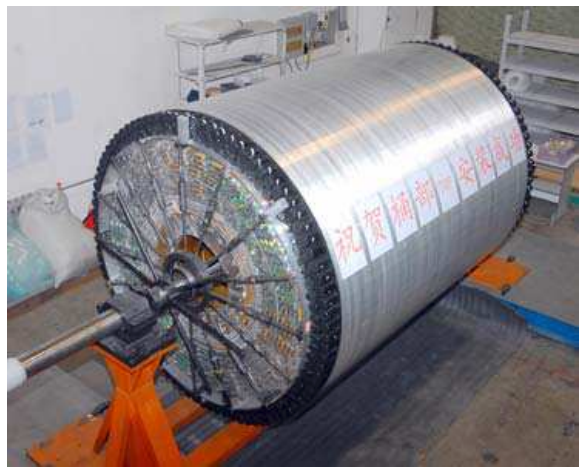}  
  \caption{\label{tof}Two layers of barrel TOF counters taped to the
    outside of the MDC before insertion into the electromagnetic calorimeter.
    The radius of the TOF counters is 870 mm.}
 \end{figure}

The TOF system, which is crucial for particle identification, is an
array of plastic scintillation counters, that measure the relative
arrival times of particles coming from electron-positron annihilations
generated by the BEPCII Collider.  It is composed of a barrel array
with 88 scintillation counters in each of two layers, and endcap (EC)
arrays with 48 fan shaped scintillators in each.  The barrel counters
are Bicron 408 scintillator of 2300 mm length and with a trapeziodal
cross section.  The endcap scintillators are Bicron 404. All TOF
counters are read out using Hamamatsu R5942 fine mesh phototubes, two
phototubes on the barrel counters (one on each end) and one on the
endcap counters.  Figure~\ref{tof} shows the barrel TOF counters
mounted on the outside of the MDC before installation into the
electromagnetic calorimeter.  The intrinsic time resolution for 1
GeV/$c$ muons with one TOF layer is less than 90 ps, which has been
confirmed using these counters and prototype BES3
electronics~\cite{wu} and using cosmic rays through the final TOF
array and read out system.
  

\section{Light source}

The light source is a PicoQuant PDL 800-B pulsed laser driver with a
LDH-P-C-440M laser diode head.  The wavelength of this head is 442 -
443 nm, and the pulse width for low power operation is 50 ps
FWHM and for high power approximately 500 ps FWHM.  Critical here is
not the pulse width but the peak power.  For the maximum power setting
of the laser driver, the peak power is 2.6 W, which provides enough
photons for our purposes. The driver can operate at 40, 20, 10, 5 or
2.5 MHz and can also be externally triggered, making the unit
convenient for use in the BES3 calibration environment.

The advantages of the laser diode compared, for instance, to a
nitrogen-dye laser used in previous systems~\cite{topaz} are ease of
operation, minimal maintenance, stability, and long lifetime.  The
laser head has Peltier cooling, and the power stability over a 12 hour
period is 1 \% RMS.

The laser head emits a collimated elliptical beam of approximately 1.5
mm $\times$ 3.5 mm and with divergence angles of 0.32 mrad parallel to
and 0.11 mrad perpendicular to the long axis of the beam.  The beam is
90 \% linearly polarized, perpendicular to the long axis of the
elliptical beam.

\section{Optical fiber bundles}

\begin{figure}  \centering
   \includegraphics*[width=0.47\pixwidth]{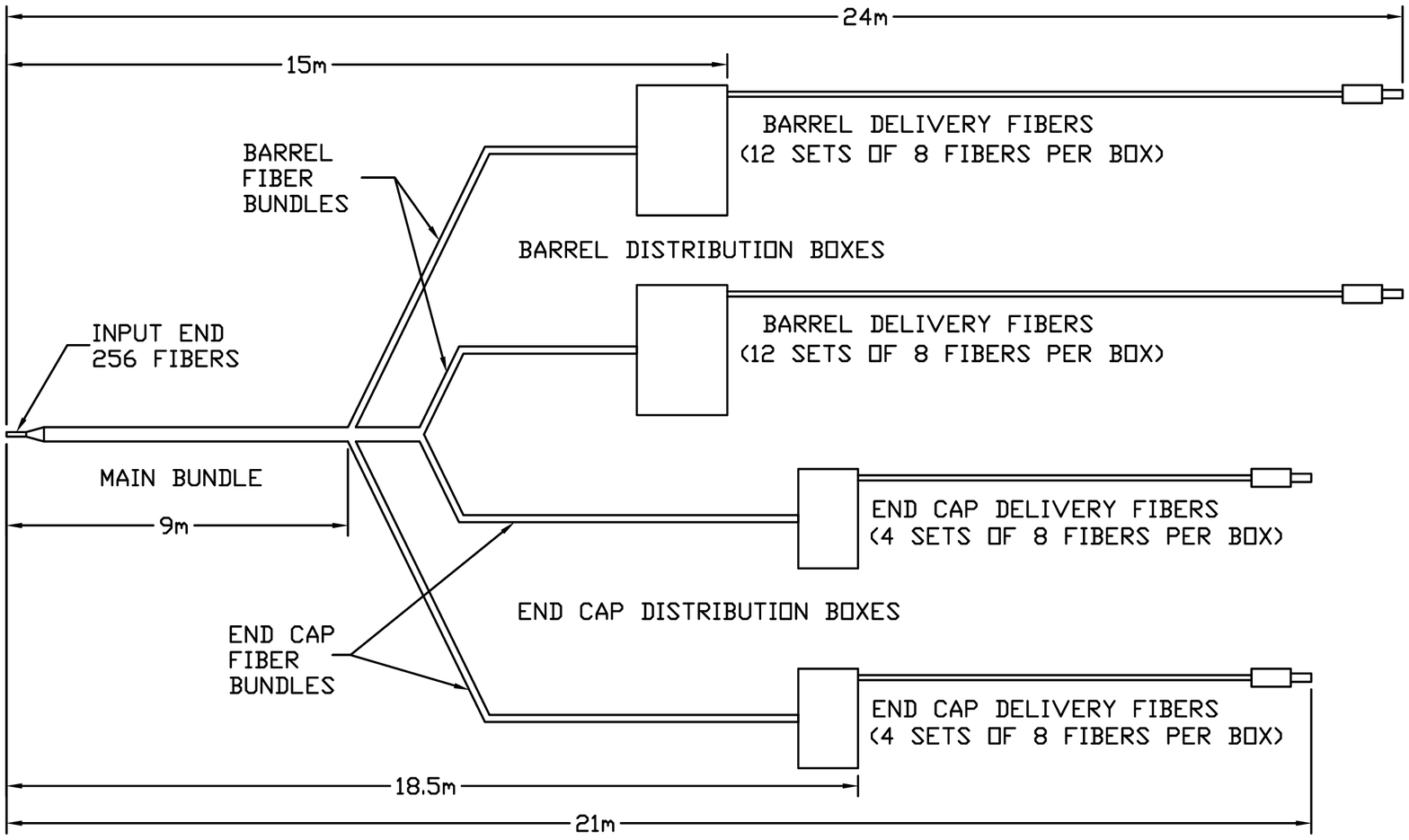}  
  \caption{\label{bundle1} Optical-fiber bundle schematic.  For
  simplicity, only one fiber/ferrule is shown from each distribution
  box.  }
 \end{figure}
 
Light from the laser diode is injected into one of two optical fiber
bundles, one for each end of the BES3 detector.  Each bundle
contains 256 optical fibers, including spares.  The optical fibers are
multimode, stepped-index, low UV attenuation (about 3 dB/km at 440 nm)
fibers, made from high OH$^-$ silica for both the core and cladding, and
they have a 100 micron core diameter, a 110 micron cladding diameter,
and a 125 micron polymide buffer diameter.  The numerical aperture is
0.22.
 
At the input end of these bundles, the optical fibers without
jacketing are closely packed into a circular cross section, which is
approximately 2.25 mm in diameter. The ratio of the core diameter to
the buffer diameter is large enough that the light coupling into the
fibers is reasonably efficient.  A 6.35 mm outer diameter by 5.0 cm
long custom ferrule here provides good concentricity between
the core bundle and its outside diameter and is mechanically
substantial enough to clamp to.  The ferrule is filled with epoxy, and
the optical fiber ends are polished.

At the back of the ferrule, the fibers are contained in a 20 cm long
PVC Monocoil tube that connects to a 7.5 cm long stainless steel
ferrule.  Inside this ferrule the 256 fibers transition into 32 PVC
Furcation Tubing jacketed cables with 8 fibers per cable.
At a distance of 9.0 m from the input ferrule, the 32 cable main bundle
divides into four sub-bundles that terminate at the fan-out
locations, as shown in Fig.~\ref{bundle1}. 
Of the four sub-bundles, two are used to illuminate one end of the
barrel TOF counters.  These sub-bundles, 6.0 m in length, have 96
fibers each contained in 12 PVC Furcation tube jacketed cables that
terminate at 8-fiber MTP connectors connected to optical-fiber
couplers inside a distribution box. The 12 cables are captured in a
mounting flange at the entrance to the box, and strain relief for the
fibers is provided. The distribution boxes allow a transition to
easily replaceable delivery optical fiber cables in the region of the
detector where the cables are most likely to be broken.

\begin{figure}  \centering
   \includegraphics*[width=0.47\pixwidth]{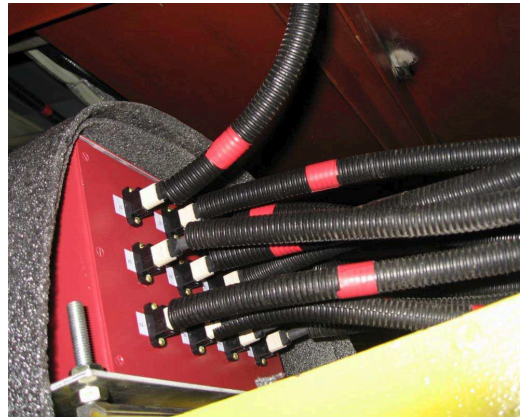}  
  \caption{\label{barrel_box} Barrel distribution box. Twelve delivery
  fiber cables in protective tubing connect to the
  MTP optical fiber couplers. Each cable has 8 fibers.}
 \end{figure}

Delivery fiber cables with 8 optical fibers and an MTP connector
mate on the outside of the distribution box, as shown in
Fig.~\ref{barrel_box}.  The cables are 9.0 m in length, but split
into individual fiber cables after 5.0 m.  These cables are jacketed
with PVC Furcation Tubing and terminate in a custom ferrule (described
below) that delivers light to the TOF counters. Mechanical strain
relief is provided wherever main bundles divide into sub-bundles
and/or individual optical fiber cables.

\begin{figure}  \centering
   \includegraphics*[width=0.47\pixwidth]{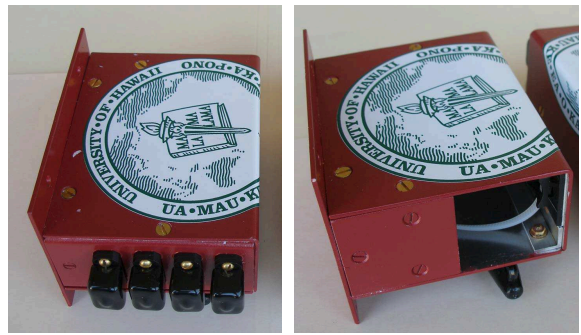}  
  \caption{\label{EC_boxes} Endcap distribution box. The left picture
    shows the four MTP connectors covered with protective caps, and
    the right picture shows the entrance and cable strap for the four
    distribution fibers from the endcap TOF counters.  }
 \end{figure}

The other two sub-bundles are used to illuminate one end of the endcap
TOF counters.  These bundles, 9.5 m in length, have 32 fibers each
divided among four cables of 8 fibers that terminate at 8-fiber MTP
connectors. These MTP connectors connect to optical-fiber couplers on
the outside of endcap distribution boxes, one of which is shown in
Fig.~\ref{EC_boxes}.

Four delivery fiber cables with 8 optical fibers and an MTP connector
connect to the couplers on the inside of the box.  The cables
are 2.5 m in length, but split into individual fiber cables after 1.0
m.  These cables terminate in custom ferrules at the end that
deliver light to the endcap TOF counters.

\begin{figure}  \centering
\includegraphics*[width=0.45\pixwidth]{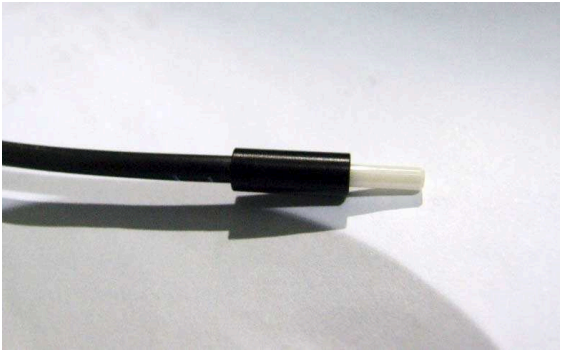}
\caption{\label{ferrule1} Delivery fiber ferrule. }
\end{figure}

\begin{figure}  \centering
\includegraphics*[width=0.45\pixwidth]{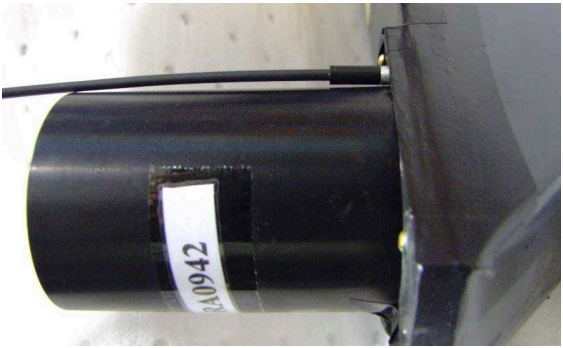}
\caption{\label{ferrule2} Ferrule
    inserted into PMT mounting flange on endcap TOF counter.}
\end{figure}

If access is required to the central detector, for instance to fix
barrel TOF phototubes or electronics or MDC electronics, the endcaps
need to be opened.  This requires that the endcap optical fiber
bundles are able to be reliably disconnected and reconnected.  The two
endcap distribution boxes mount on the back of the endcap.  The size
of each box is limited to 40 mm x 80 mm x 100 mm (see
Fig.~\ref{EC_boxes}).  Both of these constraints necessitate the use of
compact multi-optical-fiber connectors here.

Spare optical fibers were included in the design of the main bundles,
in the event that breakage occurs during fabrication or installation.
There are a total of 32 spare
optical fibers out of 256 total optical fibers (= 12.5\% spares).
Spare delivery fiber cables for both the barrel and endcap distribution
boxes were also purchased.

Figure~\ref{ferrule1} shows the ferrule that is common for all TOF
counters. The ferrule tip is zirconia and the length and diameter of
the tip are $7.95^{+0.05}_{-0.15}$ mm and $2.49 \pm 0.01$ mm,
respectively.  The 1.27 cm long transition tube is Delrin with a
diameter of 4.75 mm.  The ends of the ferrules have a standard
fiberoptic polish.  The ferrules are inserted into holes within the
mounting flanges of the phototubes, as shown in Fig.~\ref{ferrule2},
and are glued in place with a silicon adhesive.  For the endcap
counter, one of which is shown in Fig~\ref{ferrule2}, the light is
injected on the top of the counter, and some reflects off the bottom
of the counter up into the PMT.  For the barrel counters, the light is
also injected into a hole in the mounting flange of each PMT and
travels the length of the scintillator into the other PMT.

\section{Laser box}

\begin{figure}  \centering
   \includegraphics*[width=0.47\pixwidth]{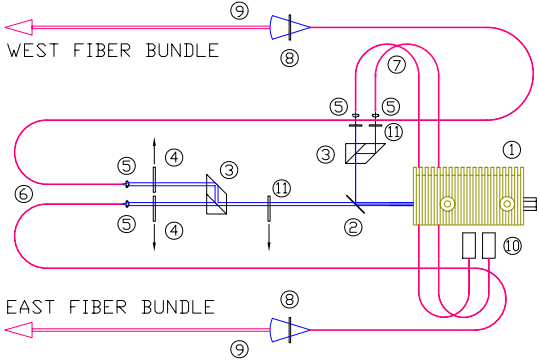}  
  \caption{\label{laser_box} Schematic of laser box components. (1)
    Laser diode head, (2) high transmissivity beam splitter, (3)
    lateral displacement beam splitters, (4) rotary solenoids, (5)
    Thorlabs 11 mm collimation optics packages, (6 and 7) 550 micron
    dismeter fibers, (8) diffusers, (9) input ends of fiber bundles,
    (10) reference PMTs, and (11) neutral density filters. 
    }
 \end{figure}

The laser box houses the laser diode head, two compact Hamamatsu
R7400U series reference photo multipliers (PMTs), and components to
couple the 1.5 mm $\times$ 3.5 mm elliptical laser beam to the 256
fibers at the input end of the fiber optics bundles and provide nearly
uniform illumination to them.  The laser box schematic is shown in
Fig.~\ref{laser_box}.  The laser diode head (1) is on the right side of
the box and shines to the left.  The first optical element (2) is a custom
2.5 cm diameter by 0.10 cm thick, high transmissivity beam splitter
with an anti-reflective coating on the incident side for 440 nm light.
Approximately 5\% of the light is split off for the reference PMTs.
Both the undeflected (fiber bundles) and deflected (reference PMTs)
beams are next incident on lateral displacement prismatic beam
splitters (3) (Edmund Optics NT47-189) that each produce two beams with
nearly equal intensity but separated by 20 mm.  In the forward
direction (fiber bundle beams), two rotary solenoids (4) control which
fiber bundle will be illuminated by blocking the undesired beam.  The
four split beams are next incident on Thorlabs 11 mm collimation
optics packages (5) (F230FC-A) that couple the light into 550 micron
diameter fibers (6 and 7), which mix the light and carry it to its next
destination.  For the fiber bundle beams, the fibers are 4 m long; for
the reference PMT beams they are 2 m long.

The light from the two fiber bundle beams is directed next to 2.5 cm
diameter diffusers (8) (Thorlabs ED1-C20) with $20^0$ divergence angles
that are located 20 mm in front of the input ends of the fiber bundles (9)
and 20 mm from the end of the 550 micron fiber.  The other two fibers
couple to the reference PMTs (10).

A third solenoid (11) under computer control and located after the high
transmissivity beam splitter, but before the lateral displacement
prismatic beam splitter, allows a neutral density filter to be inserted
into the undeflected beam to be able to balance the amount of light
going to the barrel and endcap TOF counters.  Neutral density filters (11)
are also placed in front of the reference PMTs to reduce the amount of
light that they receive.  Moveable stages are used at all key points
in order to allow precise alignment of optical components.

\section{TOF monitor control electronics}

\begin{figure}  \centering
   \includegraphics*[width=0.48\pixwidth]{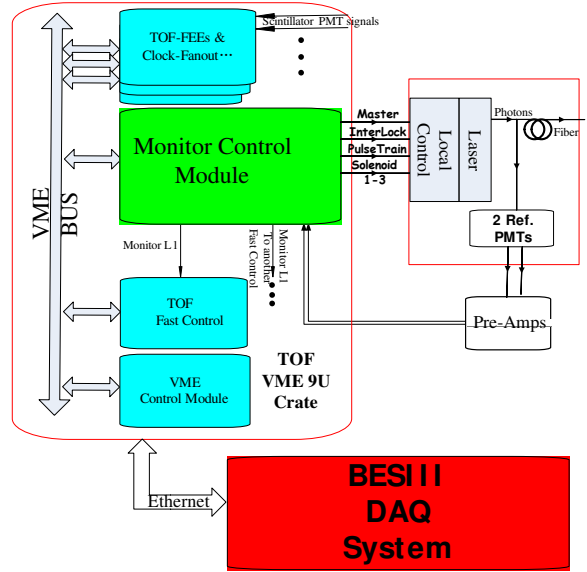}  
  \caption{\label{USTC_module} TOF monitor control electronics.
    }
 \end{figure}

The laser monitoring system interfaces with the BES3 data monitoring
program and the data acquisition (DAQ) system using a VME TOF monitor
control module, as shown in Fig.~\ref{USTC_module}. This module
enables the laser interlock, sends trigger pulses to the laser driver,
controls the three solenoids, and digitizes the time and charge of the
pulses from the two reference PMTs using the same HPTDC readout scheme
as is used for the TOF counter PMTs. The desired number of triggers,
trigger rate, and solenoid status are under the control of the TOF
monitoring program.  After each laser pulse, the VME module sends a
monitor level 1 trigger to the fast readout control.

A second control interface was designed to allow manual operation of the
laser monitor system in ``stand alone'' mode.  It has toggle switches
to control the solenoids, the laser interlock, and to send trigger
pulses to the laser driver, as well as potentiometers to adjust the reference
PMT control voltages.  This system also acts, when set to ``remote
mode'', as an interface between the TOF monitor control electronics and the
laser box components. In either mode, it supplies the actual power to
the solenoids, the interlock enable voltage, the trigger pulses, and
the necessary voltages for the reference PMTs.  The heart of the second
control interface is a ML403 Xilinx board that houses a Virtex 4 FPGA.

\section{Optical fiber bundle tests \label{bundle}}

\begin{figure}  \centering
   \includegraphics*[width=0.48\pixwidth]{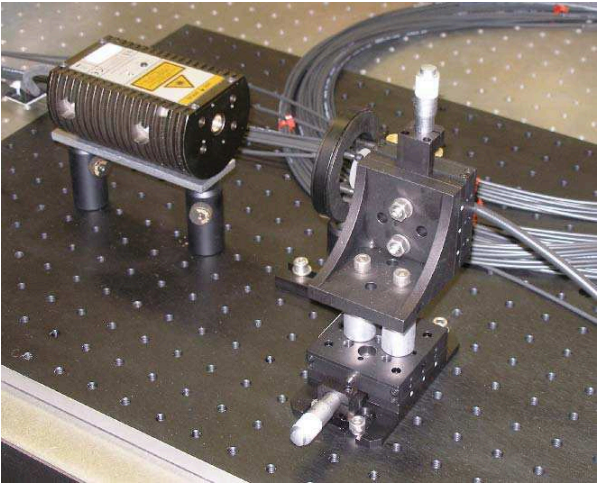}  
  \caption{\label{fiber_measurements} Fiber bundle test setup.  The
 PicoQuant laser diode head illuminates the common end of a fiber
 bundle through a diffuser that provides nearly uniform illumination
 at the input end.}
 \end{figure}

Before installation, the fiber bundles were tested for light
transmission uniformity and length uniformity.  The test setup is
shown in Fig.~\ref{fiber_measurements}.  The PicoQuant laser diode
head operating at 1 kHz illuminates the common end of a fiber bundle
through a diffuser that provides nearly uniform illumination at the
input end.  One fiber is chosen as the reference, and the differences
between the arrival times of the laser pulses from the reference fiber
and all other fibers are determined.  The measurements are made with
the two Hamamatsu R7400U reference PMTs, and the pulse times and charges are
measured with a V1290A, 32 channel, 25 ps CAEN TDC and a V965A, 8
channel dual range, CAEN QDC.

\begin{figure} \centering
\hspace{-15 pt}
\rotatebox{-90}
{\includegraphics*[width=0.38\pixwidth]{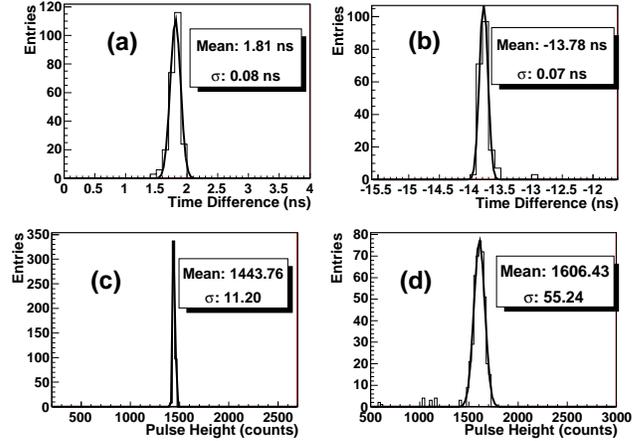}}  
\caption{\label{fiber_measurements2} Fiber test results: 
  Distributions of time differences (ns) between reference fiber and all
  other (a) barrel fibers and (b) endcap fibers. (c) Q distribution
  (in counts) 
  of reference fiber. (d) Q distribution (in counts) of all other fibers. }
\end{figure}

Fig.~\ref{fiber_measurements2} shows the time differences with the
reference fiber for (a) barrel fibers and (b) endcap fibers.
Important here is the resolution, which is $\sigma = 83$ ps for the
barrel fibers and $\sigma = 67$ ps for the endcap fibers.  These
include the read out resolution plus contributions due to the
variation of the fiber's lengths. Assuming the length variation is
dominant, the 67 ps time resolution for the endcap corresponds to a
length variation of $\sigma =1.3$ cm.
Figure~\ref{fiber_measurements2} (c) and
Fig~\ref{fiber_measurements2} (d) show the charge distributions of the
reference PMT and all other PMTs, respectively.  The resolution of the
reference PMT is 0.8\%, which includes the resolution of the charge
readout plus the variation of the laser diode intensity and
demonstrates the great stability of this device.  The charge
resolution of all other fibers is 4.0\%, which has contributions from
the charge readout and the pulse-to-pulse variation of the laser diode
intensity, as well as the uniformity of the laser beam over the input
end of the laser bundle and the transmission uniformity of the fibers.
  
\section{Laser bundle illumination uniformity}

\begin{figure}  \centering
\rotatebox{-90}
{   \includegraphics*[width=0.33\pixwidth]{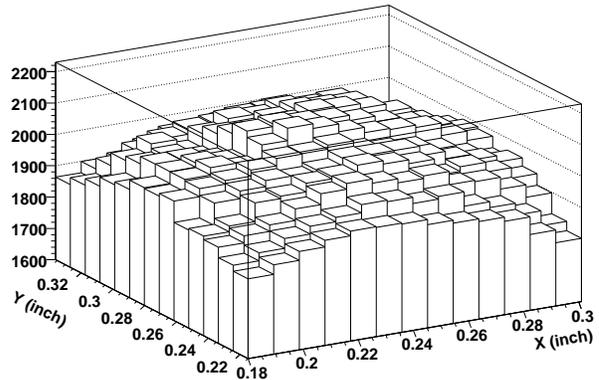}}  
  \caption{\label{scan} Result of scan with a single fiber at the
  location of the
  input end of one bundle. Shown is the charge read out
  as a function of position over a region of approximately 3.0 mm x
  3.0 mm. Note that the vertical scale is zero suppressed.}
 \end{figure}

The illumination uniformity at the input end of the fiber bundle was
measured by scanning a single fiber both horizontally and vertically
across the light beam exiting from one of the diffusers in the laser
box, shown in Fig.~\ref{laser_box}.  The result is shown in
Fig.~\ref{scan}.  The percentage RMS variation over a region of 3.0 mm
x 3.0 mm is 3.1\%.  This contributes a large component to the charge
resolution of 4\% that was measured for all other fibers in
Fig.~\ref{fiber_measurements2} (d).

\section{Performance}

\begin{figure}  \centering
\rotatebox{-90}
{   \includegraphics*[width=0.32\pixwidth]{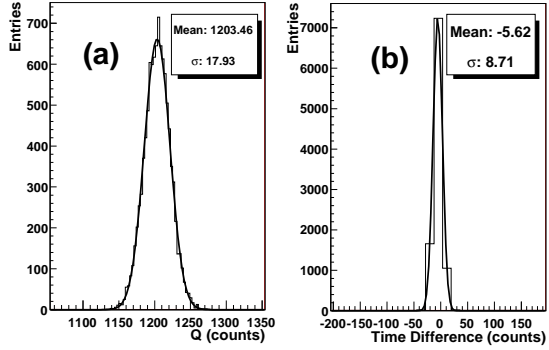}  }
  \caption{\label{refpmt} (a) The Q distribution for reference PMT
1; the resolution is about 18 counts or 1.5\%. 
(b) The time difference between the two reference
PMTs; the resolution is 8.7 counts or about 26 ps. }
 \end{figure}

Here results of the TOF monitoring system, readout using the full DAQ
system, are shown after installation in the BES3 detector.  The
system generates 10,000 laser pulses at a 1 kHz repetition rate to
either the west end or east end of the BES3 detector.  Shown in
Fig.~\ref{refpmt} (a) is the Q distribution for reference PMT 1; the
resolution is about 18 counts or 1.5\%. Shown in Fig.~\ref{refpmt} (b)
is the time difference between the two reference PMTs; the resolution
is 8.7 counts or about 26 ps, which is consistent with the
expected resolution of the TOF readout system using HPTDC chips.

\begin{figure}  \centering
\rotatebox{-90}
{   \includegraphics*[width=0.32\pixwidth]{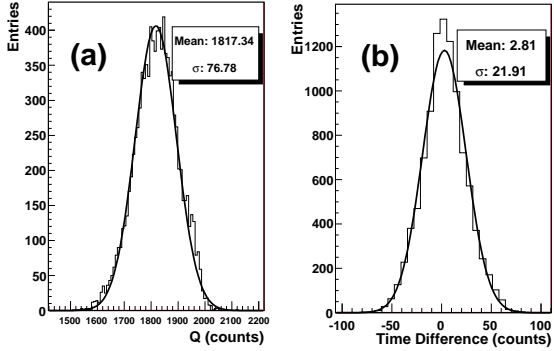}  }
  \caption{\label{tofpmt} Distributions for a typical TOF PMT. (a) Q
distribution; the resolution is about 77 counts or 4.2\%. (b) The
time difference with a reference PMT; the time resolution is 22
counts or about 66 ps.}
 \end{figure}

Figure~\ref{tofpmt} shows distributions for a typical TOF PMT
of Q and the time difference with a reference PMT after the
``time-walk'' correction.  The Q resolution
is about 77 counts or 4.2\%, and the time resolution is 22 counts
or about 66 ps.  The ``time-walk'' correction accounts for the
dependence of the measured time on the measured charge of the pulse.
The correction is of the form:
$$T_{corr} = A + BQ^{-\frac{1}{2}} +CQ^{-1}. $$

\begin{figure}  \centering
\rotatebox{-90}
{   \includegraphics*[width=0.34\pixwidth]{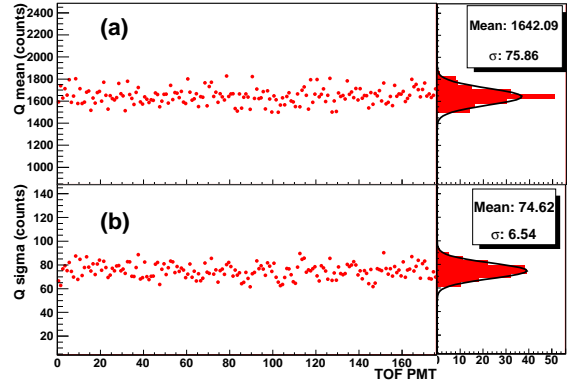}}  
  \caption{\label{qvspmt} Measured Q mean and sigma versus
  west barrel PMT number. Projections for all PMTs are shown on the right.}
 \end{figure}

\begin{figure}  \centering
\rotatebox{-90}
{   \includegraphics*[width=0.34\pixwidth]{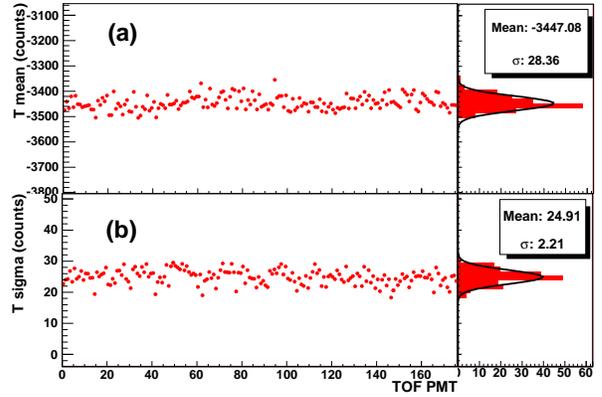} } 
  \caption{\label{dtvspmt} Time difference between TOF counter and
  reference counter versus West Barrel PMT number. Projections for all
  PMTs are shown on the right.}
 \end{figure}

Figure~\ref{qvspmt} shows the mean and sigma of Q for all west barrel
PMTs versus the PMT number.  This plot is before fine adjustment of
PMT high voltages. Figure ~\ref{dtvspmt} shows the mean and sigma of
the time differences
with a reference PMT for all west barrel PMTs.

\section{Summary}

Initial tests show that the TOF monitoring system using a PicoQuant
laser diode and custom built fiber optic bundles is very
successful. The resolutions obtained with R7400U reference PMTs are
1.5\% for Q and about 24 ps for the time difference of the two PMTs.
For a typical barrel counter PMT after the time-walk correction, the
resolutions are 4.2\% and 66 ps, respectively.  This system will be
used to test the TOF PMTs and electronics on a daily basis
and check the performance of each PMT against historical values
saved in a database.

\section{Acknowledgements}
We wish to thank all of our BES3 collaborators who helped make this
work possible. We also want to thank Hiromichi Kichimi for much
guidance in the design of this system, and Weiguo Li for useful
suggestions on this paper. This work is supported by the
BEPCII project, the CAS Knowledge Innovation Programs U-602, U-34
(IHEP), the National Natural Science Foundation of China (10405023),
and the U.S. Department of Energy under Contract No. DE-FG02-04ER41291
(U. Hawaii).

\end{document}